\documentclass[doublecol]{epl2} 

\usepackage{standalone}
\usepackage[utf8]{inputenc}
\usepackage{amsmath}
\usepackage{graphicx}
\usepackage{color}
\usepackage{comment}
\usepackage{amssymb}
\usepackage{amsmath}
\usepackage{amsthm}
\usepackage{commath}
\usepackage{float}
\usepackage{graphicx}
\usepackage{mathtools}
\usepackage{hyperref}
\usepackage{xcolor}
\usepackage{xargs}                      
\usepackage{arydshln}

\usepackage{subfigure}
\usepackage{scalerel}
\usepackage{fixltx2e} 
\usepackage{url} 

\newcommand{\Ron}{R_{\text{on}}}                     
\newcommand{\Roff}{R_{\text{off}}}                   

\title{Memristive Networks: from Graph Theory to Statistical Physics}
\author{A. Zegarac\inst{1,2,3} \and F. Caravelli\inst{4}}
\date{November 2018}
\institute{     
  \inst{1}ETH Zurich, 8092 Zurich, Switzerland\ \ \ 
  \inst{2}   London Institute for Mathematical Sciences, 35a South Street, London W1K 2XF, UK\ \ \ 
  \inst{3} Invenia Labs, 27 Parkside Place, CB1 1JF Cambridge (UK)\\
  \inst{4} Theoretical Division (T4) and Center for Nonlinear Studies,\\
Los Alamos National Laboratory, Los Alamos, New Mexico 87545, USA
}

\pacs{05.45.-a}{Nonlinear dynamics and chaos}
\pacs{05.20.-y}{Classical statistical mechanics}
\pacs{07.50.Ek}{Electronic circuits}

\abstract{
We provide an introduction to a very specific toy model of memristive networks, for which an exact differential equation for the internal memory which contains the Kirchhoff laws is known. In particular, we highlight  how the circuit topology enters the dynamics via an analysis of directed graph. We try to highlight in particular the connection between the asymptotic states of memristors and the Ising model, and the relation to the dynamics and statics of disordered systems. }

\begin{document}

\maketitle

\section{Introduction}
This perspective paper is concerned with the open questions that we find interesting in the context of a seemingly simple toy model of memristive ``endogeneous" dynamics. Some aspects, important for the understanding of the behavior of circuits of memristors, also apply to the equilibrium configuration of currents in a resistive network, as well as other systems in which Kirchhoff laws play an important role. By no means this paper is exhaustive, and for a broader overview of the topic of memristors we suggest the recent and less recent reviews \cite{reviewCarCar,Rev1,Rev2}. Our aim is to emphasize two aspects of the dynamics of memristors which characterizes the behavior of circuit with memory: the rather non-trivial connection between the underlying circuit and the non-linear dynamics of these components, and the relationship between the Physics of disordered systems and the dynamical asymptotic behavior of the circuit. In order to accomplish this task, we generalize the previously obtained equation to the case in which the disorder is present, and study the new equation.

In the late 2000s, researchers at Hewlett-Packard realized \cite{stru8,Valov,stru13} that many transition metal dioxides had the properties, initially theorized by Chua \cite{chua71,chua76} in the early '70s, of possessing an internal memory and a hysteretic behavior; certain metal oxides such as those derived from Tungsten or Titanium have the interesting property that the resistance changes noticeably as a function of time. The state of the resistance between two limiting values can be parametrized by a  parameter $w$, which is constrained between 0 and 1. We will refer to this parameter as the \emph{internal memory parameter}. For the case of titanium dioxide, the evolution of the resistance was described by the following two equations:
$$ R(w)=\Ron (1-w) +w \Roff\equiv \Ron(1+\xi w),$$
$$ \frac{\dif}{\dif t} w(t)=\alpha w- \frac{\Ron}{\beta} I(t),$$ 
initially studied for $\alpha=0$, and where $0\leq w\leq 1$, $\xi=\frac{\Roff-\Ron}{\Ron}$, and $I(t)$ is the current flowing in the device at time $t$. If $\alpha=0$, the second equation can be integrated and $w$ shown to be directly related to the charge in the conductor. Albeit this model has been revisited several times, it still serves as a prototypical model of a memory-resistor: the memristor. Also, it became clear that memory is a very common feature of nanoscale component \cite{pershin11a,diventra13a}. The interest in these components is due to the fact that memristors can serve for the purpose of neuromorphic computing \cite{indiveri}. For a more technology-oriented review of the subject, we suggest in particular the recent review \cite{reviewCarCar}.

Given the brevity of this article, 
we  will focus on the stylized facts known for the vectorial differential equation, derived in \cite{Caravelli2016rl}. The equation describes the time evolution of the internal memory of a memristive circuit (circuit made of memristors only) for the case of homogeneous memristors ($\alpha,\beta,\Ron,\Roff$ identical across the network):
\begin{equation}
\frac{d}{dt}\vec{w}(t)=\alpha\vec{w}(t)-\frac{1}{\beta} \left(I+\xi \Omega W(t)\right)^{-1} \Omega \vec S(t),
\label{eq:diffeq}
\end{equation}
where $w_i(t)$ are the internal variables constrained between $0$ and $1$, $W_{ij}(t)=w_{i}(t) \delta_{ij}$, $I$ is the identity matrix, $\Omega$ is a matrix which we will describe soon, and $S_i(t)$ is a vector of the voltage sources.  
We can see that the constants $\alpha,\beta$ and $\xi$ control the decay and reinforcement time scales and the nonlinearity in the equation respectively. In principle, we could discuss the properties of this equation without referring to where it came from, but for a deeper understanding of its properties it is useful to understand its origins. We can think of the variables $w_i$ as living on the edges of a graph,  where $\Omega_{ij}$ (which we anticipate to be a projector operator) contains the information about the topology of the graph. Each edge of the graph represents a resistive component. 

There are several limits to the applicability of eqn. (\ref{eq:diffeq}).
For instance, it has been derived only for ideal memristors, in which the time derivative of the internal parameter depends only linearly on the current, and is ideal: no parasitic capacitance or inductance are considered in the dynamics. Also, we consider only an endogenous dynamics, e.g. there are no voltage or current generators in parallel to the system (circuit) under scrutiny. This implies that for instance the interesting dynamics of \cite{PershinMaze}, in which it has been shown that memristors can be used to solve a maze, cannot be analyzed using the approach of this paper. Also, the possibility of solving the equation in full generality is an illusion, due to the constraints which make the differential equation discontinuous. What is the purpose then?

The advantages of using a toy model for analyzing a system which would be, otherwise, much more complicated are multiple. For instance, if the circuit is controlled with sinusoidal voltage and none of the memristor reaches the boundary, then the dynamics is continuous and a solution of the equation provides a solution for the evolution of each single memristor. A solution for small values of $\xi$ and $S$ when controlled with sinusoidal voltages has been provided in \cite{Caravelli2016rl}. Also, when controlled with constant voltage, the dynamics of the circuit (as we will discuss below)  is interesting enough to serve as a good toy model to the relaxation of more general circuits with memory. If for instance there are parasitic capacitance and inductance and the system is controlled with sinusoidal voltages, the dynamics of eqn. (\ref{eq:diffeq}) can still serve to analyze the dynamics for longer time-scales (where now we replace $R(t)\rightarrow Z(t)$, the admittance). Also, how do the Kirchhoff constraints affect the dynamics? Are there any hidden symmetries? How do memristors interact in the short term dynamics? Is there an emergent speed of light in the system? As we will see, these questions can be asked (and answered) with the toy model above. 

For instance, the differential equation (\ref{eq:diffeq}), written in this form, highlighlites some non-obvious symmetries of the dynamics of memristors. In fact, since we can always write $\vec S=\Omega \vec S+(I-\Omega) \vec S$, it is easy to see that we can add to $\vec S$ any vector $\tilde S=(I-\Omega) \vec k$, which will not affect the dynamics. This form of freedom arises from the Kirchhoff constraints from which the differential equation has been derived. It is due to the Kirchhoff constraints that in principle the system could have long-range interactions; this is one of the formal arguments we cover in this paper. 

Specifically, the first half of this paper focuses on the connection between $\Omega$ and the graph (the circuit), while the second part on the properties of the differential equation which might be of interest to an audience of Statistical Physicists. Albeit most of the work discussed in this paper is not novel, there are some novel points of view that we wish to share along with recent numerical simulations.

\section{From Graph theory..}
Let us first provide a simple explanation of the origin and applications of the matrix $\Omega$, as its use is not new and deeply connected to constrained flows, and thus resistive circuits.
What we state below about the graph theoretical approach to memristors is true in fact for resistor networks as well.
Let us consider a network of resistors connected \textit{in series} to voltage generators; the graph $G$  represents a circuit, and to each edge $e_i$ of the graph we can associate a pair of variables $(R_i, S_i)$, where $R_i$ is the resistance and $S_i$ the voltage.
If $R_{ij}=R_j \delta_{i j}$ is the diagonal matrix of the resistances, then it is known that \cite{bollobas2012graph,Caravelli2016rl}:
\begin{equation}
    \vec I=-A^t(A RA^t)^{-1} A \vec S,
\end{equation}
where the vector $\vec I$ is the equilibrium configuration of the currents on each edge of the graph, and $A$ is the cycle matrix of the graph, which we will define shortly. 
We note that in the case of unit resistances, we have
\begin{equation}
    \vec I=-A^t(AA^t)^{-1} A \vec S=-\Omega \vec S,
    \label{eq:unit}
\end{equation}
where we can recognize $A^t(AA^t)^{-1} A$ as a projector, which we denote by $\Omega$ and which satisfies $\Omega^2=\Omega$. It is not obvious to see this, but $\Omega$ expresses the Kirchhoff laws \cite{Caravelli2016ml} for the circuits. It is however interesting to note that $\Omega$ is a generalization of the concept of effective resistance. Let us assume for instance that voltage is applied to only one edge in the network. We denote that edge by $e_k$ and we label the vertices at its ends by $v_1$ and $v_2$.
Then we have that, since the current that flows into the resistor network must flow out, i.e., $\frac{S_{k}}{1+R_{k}}= I_k$. It is easy to see, using equation (\ref{eq:unit}), that $R_{k}$ is the definition of effective resistance between the nodes $v_1$ and $v_2$, and thus we have:
\begin{equation*}
    R_{k}=1+\frac{1}{\Omega_{k k}}.
\end{equation*}
In addition to this information, $\Omega_{kl}$ will also contain the information about the current flowing through the resistance $k$ in the network as we apply the voltage in series on the resistance $l$. In order to be more precise about the properties of $\Omega$, we provide a quick graph theoretic introduction. A (directed) graph consists of two objects: vertices and edges. Vertices can be thought of as points and edges as lines that connect some of those points. We will label vertices as $v_1, \dots, v_n$ and edges as $e_1, \dots, e_m$. Mathematically, we represent an edge starting at vertex $v_i$ and ending at vertex $v_j$ as an ordered pair $(v_i, v_j)$.
We say that a graph is \emph{planar} if it can be drawn in a plane without any of its edges intersecting. 


\subsection{Incidence matrix; $\Omega_{B^T}$}

Let $G$ be a directed graph. One way of representing $G$ is by specifying where each of its edges starts and where it ends. It is convenient to do this using a matrix. We call such a matrix an \emph{incidence matrix}.

\begin{figure}[h]
\centering
\includegraphics[scale=1]{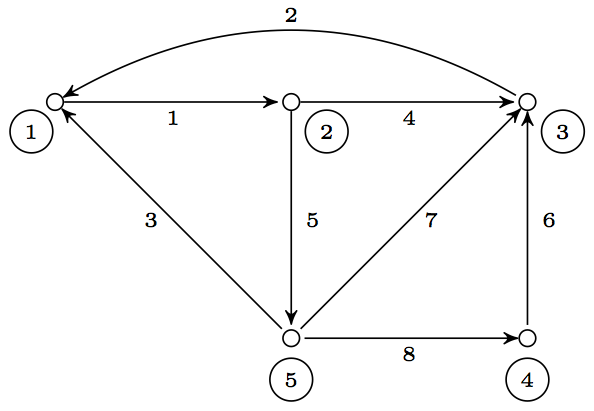}
\caption{A labelled directed planar graph.}
\label{fig:tikz-egplanargraph_labelled}
\end{figure}

As an example, consider the graph labelled as in Fig. \ref{fig:tikz-egplanargraph_labelled}.

Each column of its incidence matrix represents an edge: the first edge starts at vertex $1$ and ends at vertex $2$, so the first column of the matrix has entry $1$ in the first row and entry $-1$ in the second row. All the other entries in the first column are $0$ because none of the other vertices are a part of that edge.
By continuing this process for every edge, we get the incidence matrix $B$ of the given graph:

\begin{equation}
B =
\begin{pmatrix}
 1 &-1 &-1 & 0 & 0 & 0 & 0 & 0 \\  
-1 & 0 & 0 & 1 & 1 & 0 & 0 & 0 \\
 0 & 1 & 0 &-1 & 0 &-1 &-1 & 0 \\
 0 & 0 & 0 & 0 & 0 & 1 & 0 &-1 \\
 0 & 0 & 1 & 0 &-1 & 0 & 1 & 1 \\
\end{pmatrix}
\end{equation}

More formally, if a graph $G$ has $v$ vertices and $e$ edges, then the incidence matrix $B$ of $G$ is a $v \times e$ matrix (i.e. a matrix with $v$ rows and $e$ columns), whose entry $(i,j)$ is defined as

\begin{equation}
B_{ij} \coloneqq \begin{cases}
1 & \text{if $v_i$ is the initial vertex of the edge $e_j$,} \\
-1 & \text{if $v_i$ is the terminal vertex of the edge $e_j$,} \\
0 & \text{otherwise.}
\end{cases}
\end{equation}

If $B$ is an incidence matrix, we can define the projector operator $\Omega_{B^T}$:

\begin{equation}
\label{eq:Omega_def}
\Omega_{B^T} = B^T\left(B B^T\right)^{-1} B,\ \ \  \Omega_{B^T}^2 = \Omega_{B^T}. 
\end{equation}

If we try to compute $\Omega_{B^T}$ from the definition, we will find that the inverse $\left(B B^T\right)^{-1}$ does not exist in general. This can be solved by either considering the reduced incidence matrix (obtained by removing a row from the original incidence matrix) or by taking the pseudoinverse of the expression instead of the "regular" inverse.

\subsection{Cycle matrix; $\Omega$}

Before we can define the projector operator $\Omega$, we discuss a few more objects from graph theory. We define a \emph{walk} on a directed graph $G$ to be a sequence of vertices, say $v_1, v_2, \dots, v_n$, such that for every pair of consecutive vertices in the sequence there exists an edge that connects them, i.e. for every $i$ such that $1<i\leq n$ either $(v_{i-1}, v_i) \in E(G)$ or $(v_i, v_{i-1}) \in E(G)$, where $E(G)$ is the edge set of the graph $G$.



\begin{figure}[h]
\centering
\includegraphics[scale=.9]{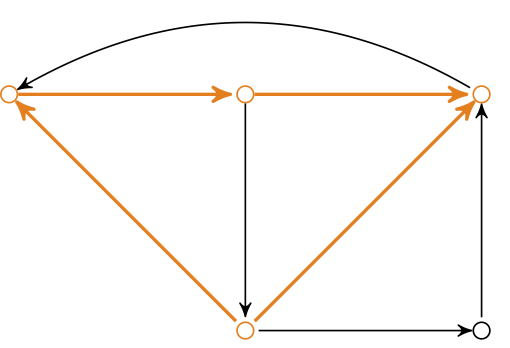}
\caption{An example of a cycle.}
\label{fig:tikz-egplanargraph_cycle}
\end{figure}

A \emph{cycle} is a walk $W = v_1 v_2 \dots v_n$ such that $l\geq 3$, $v_0 = v_n$ and the vertices $v_i$, $0<i<n$ are distinct from each other and from $v_0$. An example of a cycle is shown in the Fig. \ref{fig:tikz-egplanargraph_cycle}.

\begin{figure*}[h]
\centering
\includegraphics[width = .85 \textwidth]{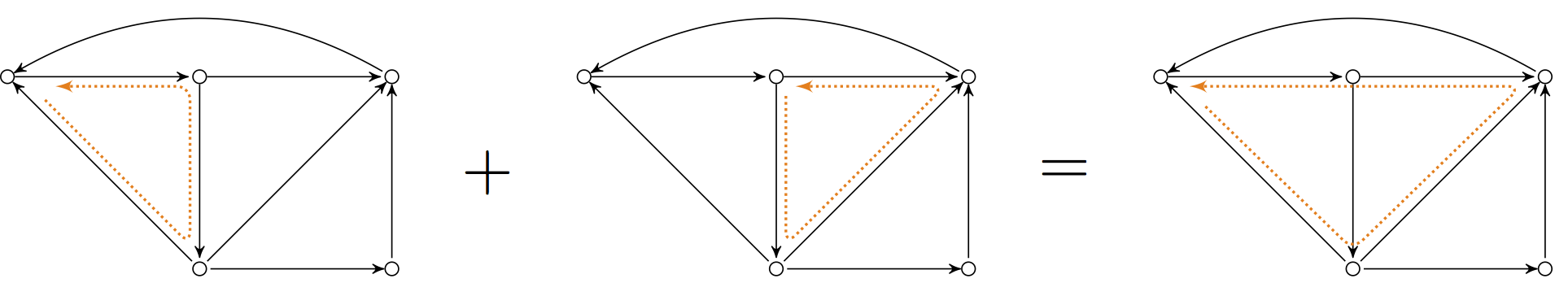}
\caption{Addition of cycles.}
\label{fig:tikz-egplanargraph_cycle_add}
\end{figure*}

The space spanned by the edges has a structure of a vector space. A \emph{cycle space} is the subset of edge space that is spanned by all cycles of a graph. A graphical representation of this fact is shown in Fig. \ref{fig:tikz-egplanargraph_cycle_add}.
A cycle matrix $A$ is a matrix whose columns form a basis of the cycle space. For example, if $\{\vec c_1, \dots, \vec c_n\}$ is a set of column vectors that form a basis of the cycle space, then the cycle matrix is    $A = (\vec c_1, \dots, \vec c_n)$.
Finally, the \emph{projector operator $\Omega$ on the cycle space of the graph} is defined to be $\Omega = A(A^T A)^{-1}A^T $. The following useful identity connects the projector operator $\Omega$ (based on the matrix $A$) to the projector operator $\Omega_{B^T}$ (based on the incidence matrix $B$) 
$\Omega = I - \Omega_{B^T}$.


\subsection{Locality} 
One question which arises immediately is: given the fact that the Kirchhoff constraints introduce some sort of non-locality between resistors (at equilibrium) and memristors (out-of-equilibrium), it is worth mentioning a few results about the matrix $\Omega$. As we will see, this problem is connected to the graph embedding problem as well. Let us first focus on planar graphs. In \cite{Caravelli2017b} the following bound on locality of interactions for planar graphs was proved:
\begin{equation}
|\Omega_{i,j}| \leq e^{-z \text{d}(i,j) + \tilde \rho}.
\end{equation}
For the purpose of this paper, we can think of $z$, $\tilde \rho$ as constants and of $\text{d}(i,j)$ as the distance between edges $i$ and $j$. The full derivation can be found in \cite{Caravelli2017b}. Here we will focus on one of the key parts of the calculation and provide a no-go theorem for the generalization of the argument for arbitrary non-planar graphs.
Finding an analytic expression for a quantity which involves an inverse of a potentially large matrix is a non-trivial problem. To overcome this in the case of $\Omega = A(A^T A)^{-1}A^T$, it was noticed that the expression for $\Omega$ simplifies if the matrix $A$ is orthonormalised first. If we denote by $\tilde A$ the orthonormalised matrix, we get $\Omega = \tilde A \tilde A^{-1}$, which is a matrix product.
The difficulty is now in calculating the orthonormalised matrix $\tilde A$ from $A$. We proceed by using a non-algorithmic expression for the Gram-Schmidt process: the $p$-th column, $A_p$, of the orthonormalised matrix $\tilde A$ is given by:

\begin{equation}
\tilde A_p =
\det \begin{pmatrix}
\langle \vec A_1, \vec A_1 \rangle & \langle \vec A_1, \vec A_2 \rangle & \dots & \langle \vec A_n, \vec A_n \rangle & \vec A_1\\
\vdots & \vdots & \ddots & \vdots & \vdots\\
\langle \vec A_n, \vec A_1 \rangle & \langle \vec A_n, \vec A_2 \rangle & \dots & \langle \vec A_n, \vec A_n \rangle & \vec A_n\\
\end{pmatrix}.
\end{equation}

We will now show how the inner products $\langle A_i, A_j \rangle$ can be expressed in terms of the adjacency matrix.

\begin{figure}[h]
\centering
\includegraphics[scale=1]{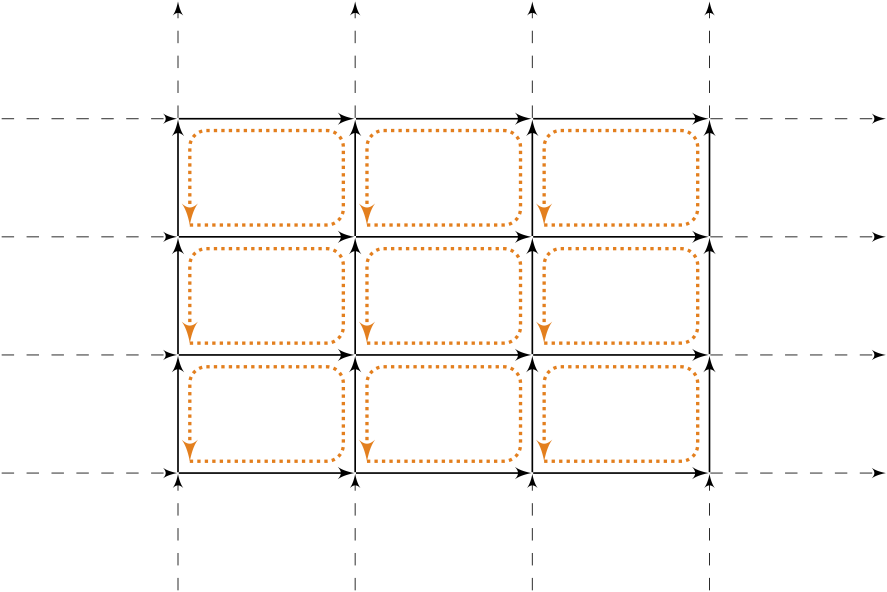}
\caption{A choice of a basis on the grid graph $G$.}
\label{fig:tikz-gridGraph3allcycles}
\end{figure}

In the case of a grid graph $G$, we perform the following steps:

\begin{itemize}
    \item First, we pick a basis of the cycle space of $G$ as in Fig. \ref{fig:tikz-gridGraph3allcycles}.
    \item Then we denote by $G'$ the graph that has a vertex for each basis cycle of $G$ and an edge between two vertices if the corresponding cycles in $G$ are adjacent.
    \item Finally, we express the inner products as

\begin{equation}
\langle A_i, A_j \rangle =
\begin{cases}
M_{i,j}, \ &\text{if $i = j$} \\
|C_{i}|, \ &\text{if $i = j$},
\end{cases}
\end{equation}
where $M_{i,j}$ is the adjacency matrix of $G'$ and $|{C_{i}}|$
is the length of the cycle corresponding to the $i$-th vertex of $G'$.
\end{itemize}
%
%
%
The last step was key as it allowed manipulations which ultimately lead to the expression for the bound in \cite{Caravelli2017b}.

Let us now focus on the case of nonplanar graphs, where the problems arise. We found that the same method cannot be applied to the nonplanar graphs, and here we provide an explanation of the reason. Physically, we would expect that for graphs that present a notion of distance (unlike for instance random graph, where the graph diamater scales as $D\approx \log(N)$ in the number of nodes $N$), a similar bound would apply. In the case of planar graphs, it was possible to choose a basis of the cycle space in such a way that the basis cycles bound the faces of the graph, greatly simplifying the proof.

A generalisation of this construction to nonplanar graphs is not obvious due to the fact that in nonplanar graphs the notion of \emph{faces} is not well-defined. In an attempt to overcome this, we defined the faces of a nonplanar graph $G$ to be the faces of an embedding of $G$ in some closed orientable surface.

\begin{figure}[h]
\centering
\includegraphics[scale=0.4]{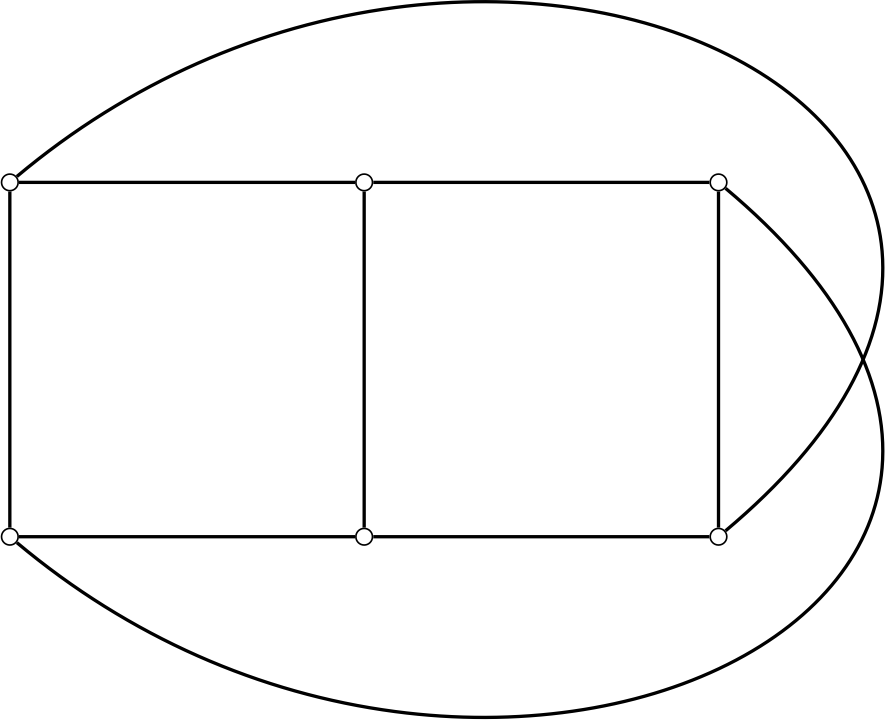}
\caption{An example of a nonplanar graph, $K_{3,3}$.}
\label{fig:tikz-K33switched}
\end{figure}

\begin{figure}[h]
\centering
\includegraphics[scale=0.3]{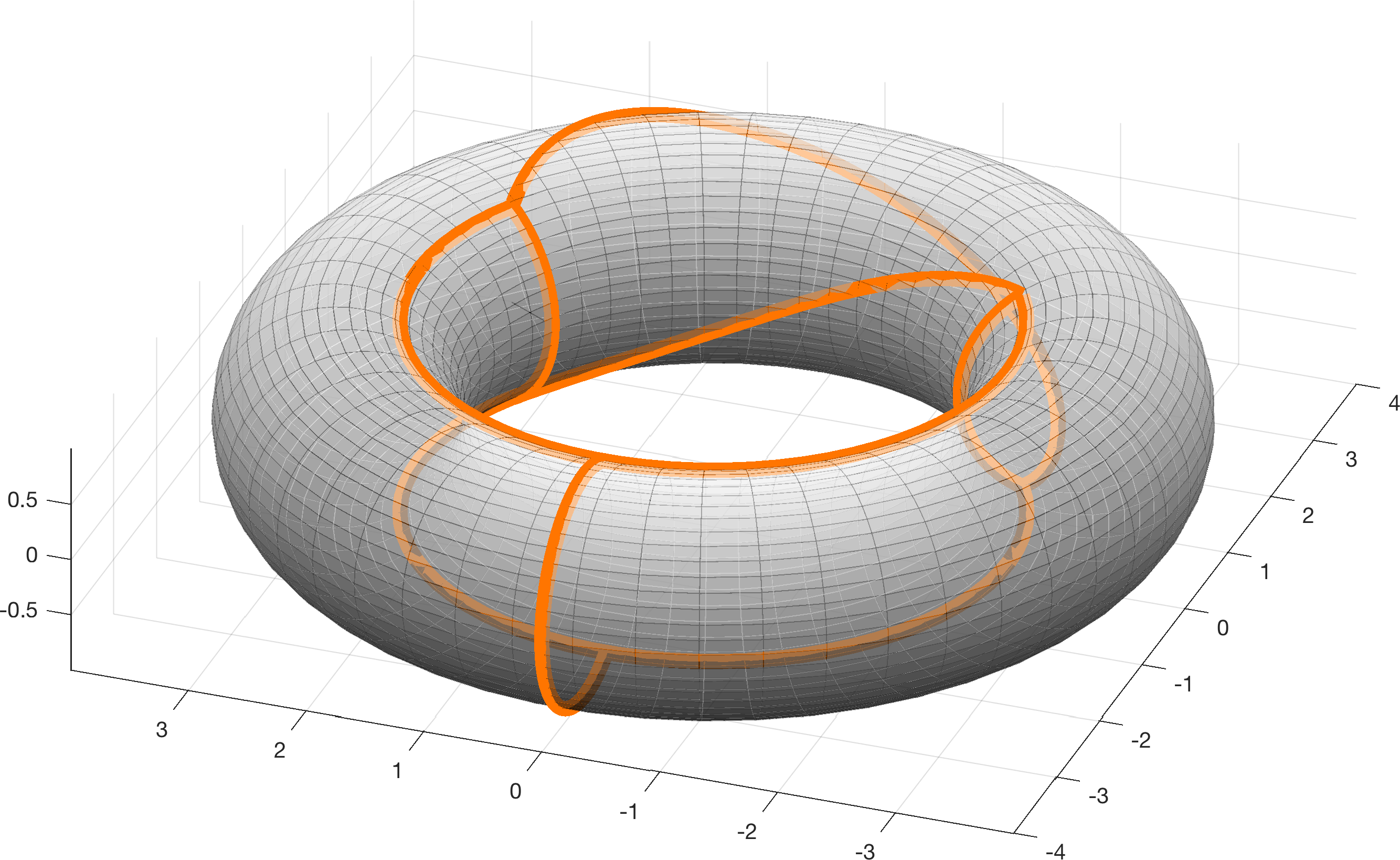}
\caption{An embedding of $K_{3,3}$ in a torus.}
\label{fig:K33torus_ppt}
\end{figure}

\begin{figure}[h]
\centering
\includegraphics[scale=0.55]{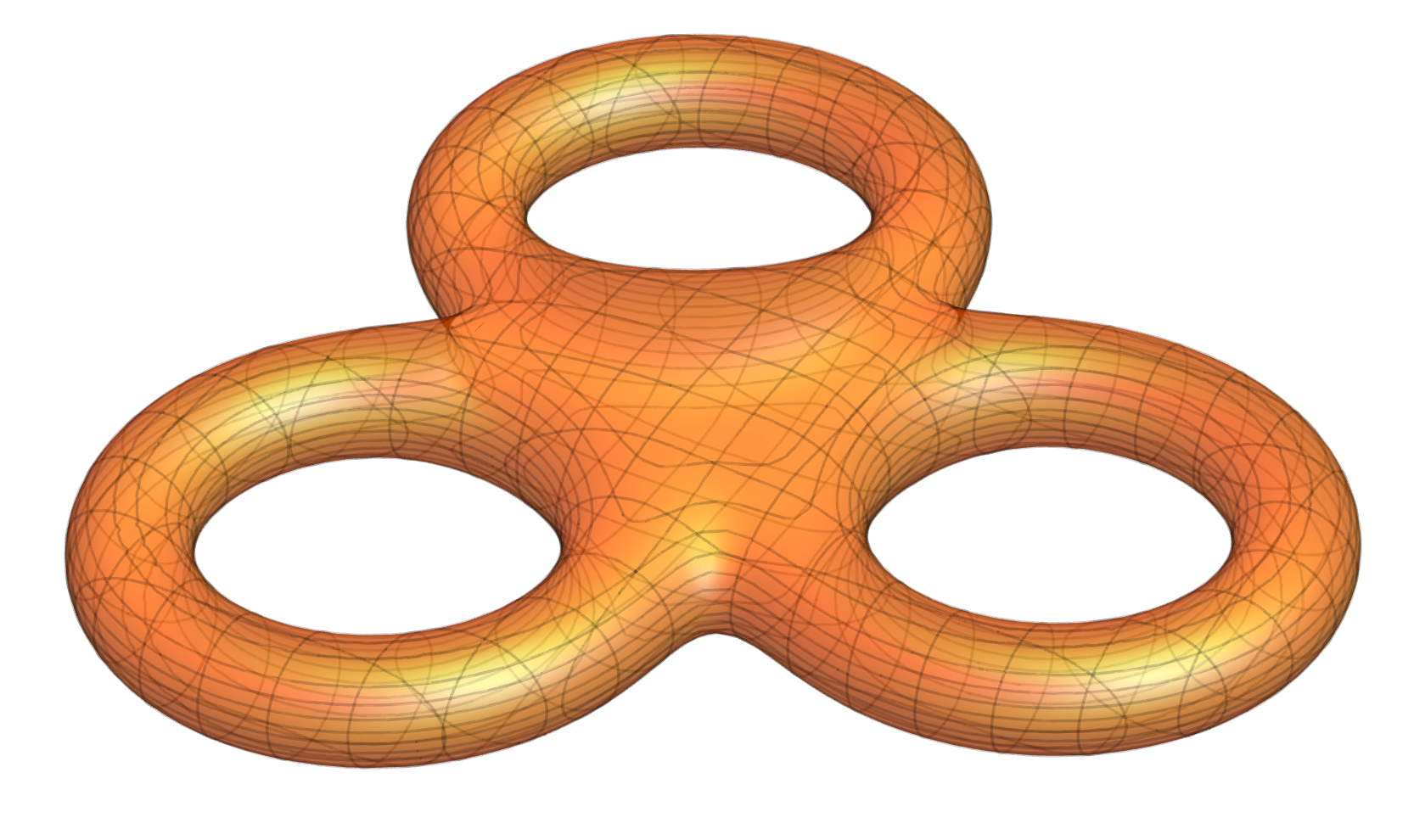}
\caption{A ``triple torus".}
\label{fig:tripletorus}
\end{figure}

We illustrate this on the simplest nonplanar graph, $K_{3,3}$, shown in Fig. \ref{fig:tikz-K33switched}.
Let $T^2$ be a torus.
An embedding of $K_{3,3}$ in $T^2$ is shown in Fig. \ref{fig:K33torus_ppt}. We thus define the faces of $K_{3,3}$ to be the connected components of $T^2 \setminus \bar K_{3,3}$, where $\bar K_{3,3}$ denotes the embedding of $K_{3,3}$ inside the torus $T^2$.
To be able to use the method from \cite{Caravelli2017b}, we have to find a basis of the cycle space consisting only of cycles that bound faces.
However, the number of elements in the basis of a cycle space of $K_{3,3}$ is 4. As can be seen in Fig. \ref{fig:K33torus_ppt} above, there are only 3 faces in the embedding of $K_{3,3}$ in $T^2$. This is a problem which we elucidate further below. Setting aside the question of well-definedness of faces, it might seem that if we embed $K_{3,3}$ in a different way or in some other closed orientable surface (for example a triple torus shown in Fig. \ref{fig:tripletorus}), we could get sufficient number of faces. We prove below that there exists no embedding with sufficient number of faces.

Let us introduce the \textit{Euler characteristic} of a graph $G$ embedded in a surface $S$ to be
\begin{equation} \label{eqn:eulerchar}
\chi_{G, S} = |V| - |E| + |F|.
\end{equation}
We want the number of faces to be greater than or equal to the dimension of the cycle space. Thus,
\begin{equation} \label{eqn:Fgeq}
|F| \geq \text{dim} \ \mathcal{C} = |E| - (|V| - 1),
\end{equation}
where $\mathcal C$ denotes the cycle space.
Substituting the equation (\ref{eqn:Fgeq}) into the expression for Euler characteristic given by equation (\ref{eqn:eulerchar}), we get
\begin{equation}
\begin{aligned}
\chi_{G, S} &= |V| - |E| + |F|\nonumber \\ 
&\geq |V| - |E| + \left(|E| - (|V| - 1)\right).\nonumber
\end{aligned}
\end{equation}
That is, we need $\chi_{G, S} \geq 1$. A graph embedded in a surface has Euler characteristic equal to the Euler characteristic of that surface.
We can also express the Euler characteristic of a surface in terms of its genus $g$ as $\chi_{G,S} = 2 - 2g$.
Hence, using the previously obtained inequality we get $ g \leq 1/2$. That is, we can only find the sufficient number of cycles that bound faces in surfaces with less that 1/2 holes. The only such closed orientable surface is a sphere (with 0 holes) and sphere is equivalent to a plane for all our purposes. Therefore we cannot use the same approach as in the planar case. This leaves us with the open question of how to generalize the bound on the interaction strength between memristors to more general non-planar graphs.

Given that the circuit enters the equation only in $\Omega$ and that it represents the Kirchhoff constraints, its study is useful The locality bound is useful for bounding also the long-term dynamics of the system. As shown for instance in \cite{Caravelli2017b}, these locality bounds imply an emergent speed of light in the system, similarly to the Lieb-Robinson bounds for quantum systems  \cite{Hasting,Schuch}:
\begin{equation}
    |\langle w_{i}(t) w_j(0)\rangle| \leq  K e^{-(d_{ij}-v_{eff} t)},
\end{equation}
where $d_{ij}$ is the Hamming distance between the memristors and $v_{eff}$ an effective speed of light. It is thus interesting to study these bounds for more general circuits that are not necessarily planar.

\section{..to Statistical Physics}
Insofar we have focused on a network of resistors via the study of the matrix $\Omega$. An initial attempt at studying the statistical properties of dynamical graphs with memory was the one of \cite{Caravelli2015}, where the emergence of scale-free networks out of the endogeneous dynamics of excitable memristor-like components was observed. 
What about a network of memristors which satisfy Kirchhoff laws? In this case the nonlinearity of the differential equation makes the analysis more complicated. The differential equation becomes vectorial as in equation (\ref{eq:diffeq}), where the variables $\vec w$ are constrained on the hypercube $[0,1]^{M}$ with $M$ being the number of memristors. It has been noted in \cite{Caravelli2016rl} that the for ``complex enough" circuits, the average parameter $\langle w \rangle=\frac{1}{N} \sum_i w_i$ relaxes slowly to the asymptotic values $w=1$ and $w=0$.

The operator $\Omega^\prime$ plays an important role in the differential equation (\ref{eq:diffeq}), as it is the only place where the graph topology enters.  In the regime of large (random) graphs, slow relaxation can be observed in the average parameter $\langle w \rangle$. A power law type of relaxation had been observed in the relaxation of $\langle w \rangle$ to the asymptotic values of $w=1$ and $w=0$ ($t^{-\rho}$ with $\rho\approx 0.92$). This feature is similar to what was observed experimentally in \cite{Avizienis,Stieg12} for atomic switch networks. What seems to be a good  parameter for the transition from fast to slow relaxation is the ratio between the number of (fundamental) circuit loops and memristors components, which is upper bounded by one.

In the case with disorder, i.e. when not all memristors have similar properties, it can be shown that a generalized differential equation for the memories still exists and is given by
\begin{equation}
\frac{\dif}{\dif t}\vec{w}(t)=A\vec{w}(t)-B^{-1} \left(I+ \Omega^\prime W(t)\right)^{-1} \Omega^\prime T^{-1} \vec S(t),
\label{eq:diffeq}
\end{equation}
where now $A_{ij}=\alpha_i \delta_{ij}$, $B_{ij}=\beta_i \delta_{ij}$ and where $T=\delta_{ij}(1+N_{ii})$ represents the \textit{disorder} among the component. The matrix $T$ characterizes the ratio $\frac{\Roff^i-\Ron^i}{\Ron^i}$ for each single memristor and $N_{ii}$ represents the disorder at fixed network topology. The matrix $T$ enters also in the modified projector operator $\Omega^\prime$. 
The operator $\Omega'$ is in fact a non-orthogonal projector, i.e. $\Omega^\prime=A^t \left( A T A^t\right)^{-1}  A  T$;
if $N_{ij}=0$, we recover the previous equation with a symmetric projector operator. This type of non-orthogonal projectors were also found relevant for the mixture of active and passive memristive components, in which $T$ can also take negative values \cite{Caravelli2016rl,riaza}.

\begin{figure}
    \centering
    \includegraphics[scale=0.24]{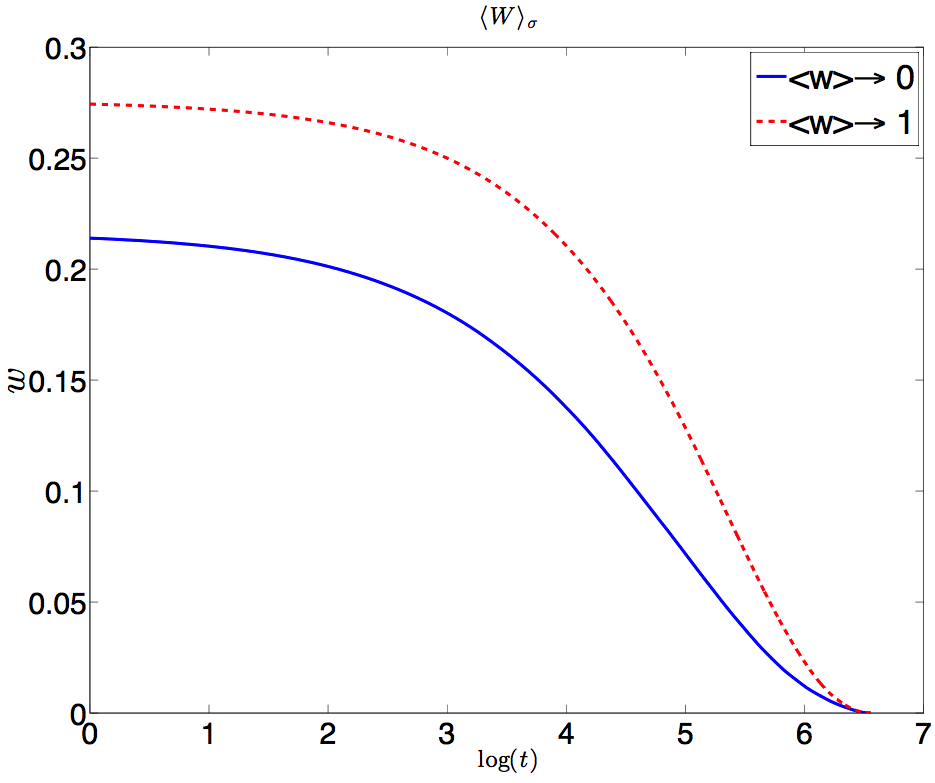}
    \caption{Average relaxation for fixed $\sigma= 0.05$, averaged over 20 simulations, on a complete graph with 50 vertices (1225 edges). The simulations were performed using $dt=0.1$, $\alpha=0.003$, $\beta=100$ and homogeneous across the system.}
    \label{fig:evolv}
\end{figure}

\begin{figure}
      \centering
      \includegraphics[scale=0.23]{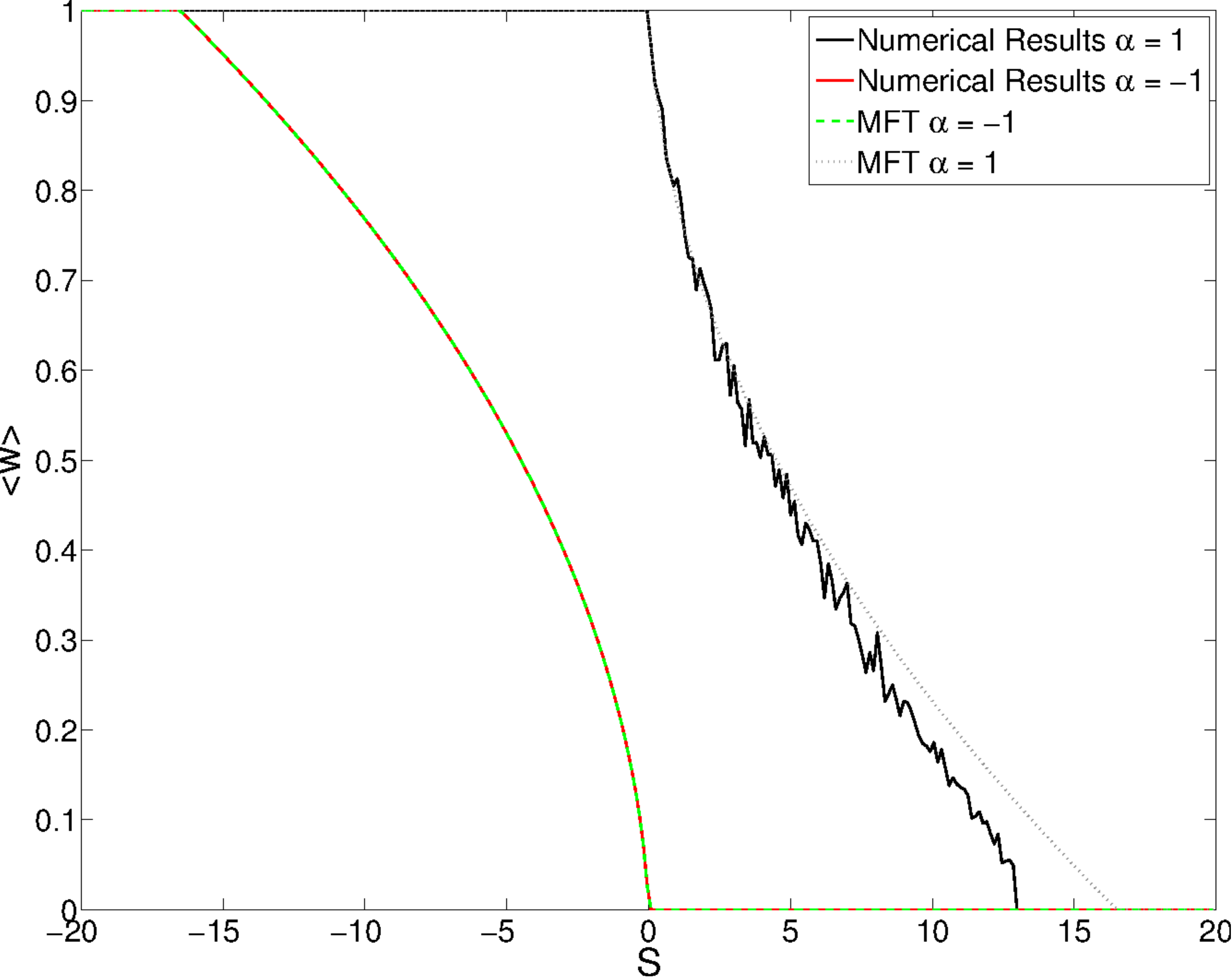}
      \caption{Mean field theory  vs numerical results for the asymptotic average parameter $\langle w \rangle$, taken from \cite{barucca}.}
      \label{fig:mft}
  \end{figure}

Recent simulations, in which we have accounted for the disorder, have shown that for larger graphs and statistics, the relaxation of the average internal memory, once we factor in the disorder, is compatible with a logarithmic one.
For instance, in Fig. \ref{fig:evolv} we plot the average parameter $\langle w\rangle$ that was obtained by averaging over different realizations of the disorder and for longer times. We see that a $\log(t)$ regime is established after an initial slow relaxation. This is one of the typical features of glassy systems which we find interesting in these rather simple systems. If we aim to use these systems for computation, glassiness in the dynamics will naturally imply a slow approach for the solution of the problem under scrutiny, for instance.
  It is interesting to note that a connection to disordered systems can be made a bit more precise by looking at the properties of the asymptotic states $w_i$ as a function of the matrix $\Omega$ even in the case with homogeneous memristors.
  It is interesting to note that the equation which describes the evolution of memristors, similarly to the case of resistors, also tries to solve another optimization problem, but this time more complicated: the QUBO, or Quadratically Unconstrained Binary Optimization and known in Physics as the ground state of the Ising model. It has been noted first in \cite{barucca} that for a model of mean field memristor interaction that a Lyapunov function, similar in spirit to the Blume-Capel model \cite{Campaetal}, exists. That model, in particular, it is shown to be exactly solvable.
  In particular, it has been suggested in \cite{barucca} that the mean field theory, similarly to the case of the Curie-Wei{\ss} model, provides a good estimate of the asymptotic value of the parameter $\langle w \rangle$, which can be considered an order parameter which parallels the mean magnetization of the Ising model. Given the fact that we are interested in the zero-temperature limit of the system, mean field theory is used at finite temperature and the temperature sent to zero at the end of the calculation. The averages are then intendended as averages over the initial conditions for the dynamics. For instance, in Fig. \ref{fig:mft} we show the order parameter $\langle w \rangle$ for the case $\alpha>0$ and $\alpha<0$ obtained via Monte Carlo and compared with the mean field theory result as a function of the mean external voltage $S$. The case $\alpha<0$ is asymptotically stable, while the case $\alpha>0$ is asymptotically unstable. Yet, the information on the position of the asymptotic fixed point of the dynamics can be used for obtaining information about the averages $\langle w \rangle$. In particular, this suggests that the mean field theory of spin glasses could be an important source of inspiration to study more general systems \cite{Parisi}.
  
For the more general case of equation (\ref{eq:diffeq}), it has been noted that the Lyapunov functional can be approximated by an Ising model (i.e. binary, rather than continuous variables), in which the exchange interaction is proportional to $\Omega_{ij}$ and with a non-zero effective external field. 
For the case of ideal memristive circuits, $\Omega$ has to be a very specific matrix based on the circuit \cite{caravelli2018}. It has been observed numerically that for random graphs, the distribution of the elements $\Omega_{ij}$ is a trimodal distribution; however, the bulk of the probability is well approximated by a zero-centered gaussian distribution with $\langle \Omega_{ij}^2 \rangle \approx \frac{1}{N}$, where $N$ is the number of memristors (or edges in the circuit). This implies that memristive networks, reinforcing the connection to the theory of mean-field spin glasses \cite{SherringtonKirkpatrick}.

From the point of view of optimization purposes however, the system of differential equations can be simulated for arbitrary $\Omega$ in principle, which suggests a heuristic optimization algorithm for QUBO type of problems. Thus, the memristive differential equation can serve as a heuristic method for tackling NP-Complete problem such as QUBO \cite{caravelli2018}. These problems are NP-Complete because there is no known algorithm that is better than exhaustive search: because of the binary nature of the variables, we necessarily have to explore all the $2^N$ possible values of the variables $\vec w$ to decide which extremum (or extrema) are better. In a certain sense, the memristive differential equation is a relaxation of the QUBO problem to continuous variables. This is the same class of the frustrated Ising model. The connection to NP-complete problems has also been observed in other memristor-based architectures \cite{Traversa2015,Traversa2014} .
The functional that the memristive networks are trying to locally maximize is the functional
\begin{equation}
M(W)=\sum_i \left(r_i-\frac{p}{2}\Sigma_{ii} \right)w_i-\frac{p}{2} \sum_{i\neq j}  w_i \Sigma_{ij}  w_j,
\end{equation}
where $r_i$ are the external fields and $\Sigma$ the exchange interaction. The mapping between the optimization of the returns above and the equivalent memristive equation is:
\begin{equation}
\begin{aligned}
\Sigma = \Omega,\ \ \  \frac{p}{2} &= \alpha \xi. \\
\frac{\alpha}{2} +\frac{\alpha\xi}{3}\Omega_{ii} -\frac{1}{\beta} \sum_j \Omega_{ij} S_j &= r_i-\frac{p}{2}\Sigma_{ii},\\
\end{aligned}
\end{equation}

We can obtain the vector $S$ through inversion of the matrix $\Sigma$, if it is invertible. Algorithmically, there is still the freedom of choosing $\xi$ and $\alpha$ given $p$, but the two limits are different in nature: $\xi\gg 1$ is the deep nonlinear regime, while $\alpha\gg 1$ is the deep diffusive regime. Nonetheless, the mapping between the asymptotic dynamics and the Ising model allows the use of inference methods for the Ising model \cite{inference}.
It is worth mentioning that  other connections to Statistical Physics of disordered systems, and in particular highlighting the importance of memristor switching for the dynamics of memristive systems have been investigated in \cite{PershinPT,sheldon}. 

\section{Conclusions} 
In the present paper we have discussed the connection between the dynamics of memristors, graph theory and the properties of certain disordered systems. This paper is meant as an invitation to the more theoretically inclined researchers. We have highlited the open problems in the analysis of a specific toy model of memristive endogenous dynamics and heuristic optimization algorithms for Quadratically Unconstrained Binary Optimization. We hope that the reader will use the toy model we have discussed in this paper as a playground for answering some precise mathematical questions regarding the interesting dynamics of memristors.

\section{Acknowledgements.}  FC acknowledges the support of NNSA for the U.S. DoE at LANL under Contract No. DE-AC52-06NA25396. AZ would like to thank Invenia Labs and the London Institute for Mathematical Sciences and the Los Alamos National Laboratory for funding and hospitality.

\clearpage
\end{document}